# Empowering the Earth system by technology:
## Using thermodynamics of the Earth system to illustrate a possible sustainable future of the planet

Axel Kleidon

**Key message:** With the use of the appropriate technology, such as photovoltaics and seawater desalination, humans have the ability to sustainably increase their production of food and energy while minimising detrimental impacts on the Earth system.

## Energy as the core problem

The future of mankind faces a multitude of challenges, with consequences that affect the planetary scale of the Earth system. Global climate change, freshwater scarcity in many parts of the world, the loss of biodiversity, and future energy and food supply are examples for some of these challenges. The multitude of challenges is overwhelming and it seems almost impossible to imagine a positive, sustainable future for humanity.  In fact, if we continue on the current path, one can easily imagine a dark outlook on the future of the planet that would seem to destroy the very conditions that are needed for a habitable planet.

What I want to describe here aims at the opposite:  To imagine a positive outlook on the future, in which human societies can sustainably grow, yet minimise their impact on the planetary environment.  Parts of this outlook were described in similar, recently published literature (Kleidon 2016, Kleidon 2019a, 2019b), yet the focus here is to put these parts into the context of envisioning a positive, sustainable future.  This future will not come automatically, but requires a substantial effort to get away from current technologies and focus on those that empower the planet.

> 

Here, I use the word „power" in its literal, physical sense of work performed in time.  Power, or the generation of free energy by natural processes mostly driven by solar radiation, drives the dynamics of the Earth system, of the biosphere as well as human societies.  Based on this view on energy generation and dissipation, I want to make three points: i) human activity consumes free energy at levels similar to other Earth system processes (Figure 1), ii) the most detrimental aspects of human activity relate to its consumption of free energy from the Earth system, rather than generating it, and iii) some forms of technology, particular photovoltaics, allow to convert solar radiation into free energy more efficiently than nature does, which provides the basis for a sustainable future.  These forms of technology provide human societies with the means to increase the power generation on Earth, make the dynamics of the planet more active, including human societies.  In other words, these technologies make human societies a producer of free energy in the Earth system.  Currently, however, most of the free energy consumed by human societies draws energy from the Earth system, thus having a detrimental effect and weakening the Earth system.

To understand this main point, I want to first focus on the central role that energy plays for humans and human societies. Energy is central for humans to be active and do things. Humans need energy to maintain their metabolisms even when they sleep and do nothing.  Humans get this energy in form of calories from the food they eat.  Yet, the majority of energy is needed for socioeconomic activity, for manufacturing, transport, heating and cooling, and so on. This energy at present comes primarily from fossil fuels, that is, dead plant material that was sealed off from the active climate system by geologic processes millions of years ago. The combustion of fossil fuels depletes this disequilibrium between hydrocarbons and atmospheric oxygen, it is not sustainable due to its final size, and the increase in greenhouse gases results in global warming.

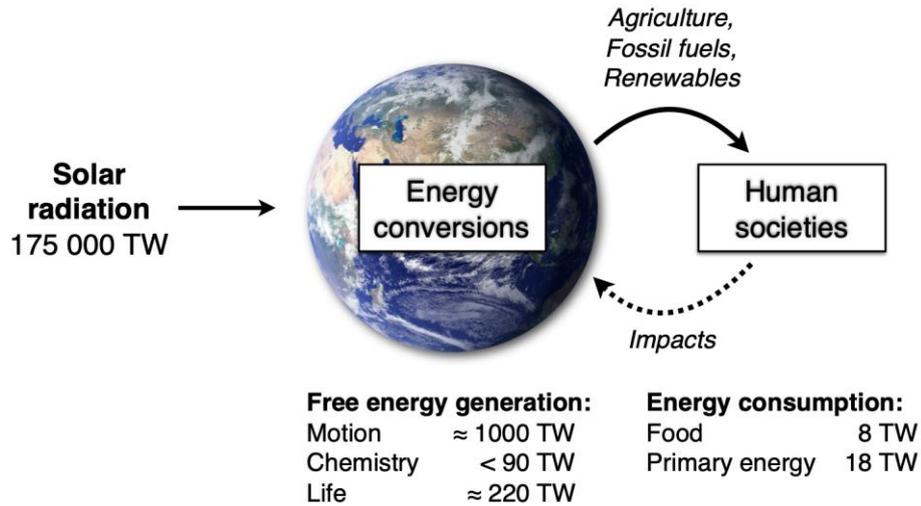

**Figure 1: Human societies consume free energy at rates that are similar to the magnitude by which Earth system processes generate physical and chemical forms of free energy. As human societies currently consume this free energy from the Earth system, this causes detrimental impacts on the planet. Image of Earth from NASA. Estimates from Kleidon (2016).**

This focus on energy makes the activity of humans comparable to the activity of other Earth system processes. I use the term "active" here to refer to this continuous need for energy to do things and to maintain dynamics. I approach this from a purely physical, or, more precisely, a thermodynamic perspective. Staying active, thermodynamically, means that processes keep producing free energy in different forms and dissipate these. A focus on this thermodynamically-based activity allows us to generalise Earth system processes, to identify evolutionary directions and basic limits, and it allows us to weave human activities that seem so different from purely natural processes into the context of the Earth system. After all, the energy used by human societies needs to come from somewhere within the Earth system, and their activities do something that ultimately alters this very same entity. By tracing energy through the Earth system to human societies and back to the Earth system, one can link these processes together and view them in the context of the functioning of the whole planetary system. We can then appreciate the basic role of humans and their technology and get a glimpse of what we can expect of the future.

**Thermodynamics sets the laws**

When we deal with Earth system processes, or processes in human societies, these are typically intimately linked with converting one form of energy into another. Atmospheric motion, for instance, is generated from converting some of the heat by absorbing solar radiation or condensation of water vapour into kinetic energy. This kinetic energy eventually turns back into heat by friction. A plant leaf converts the energy contained in sunlight into chemical energy associated with the production of carbohydrates. Plants, animals, and humans turn these carbohydrates back into heat when they respire it to maintain their metabolic activities. Likewise, technology like power plants or wind turbines are energy converters. A power plant converts the chemical energy of a fuel, such as coal, into heat during combustion, of which a fraction is converted into electric energy that is fed into an electric grid. Wind turbines similarly convert wind energy into electricity. This electric energy is used by human societies and ultimately is also converted back into heat.

These energy conversions follow a strict direction and are constrained by fundamental limits, both of which are set by thermodynamics. The laws of thermodynamics set the basic rules for energy conversions. The first law states that energy in total is conserved when it is being converted from one form into another, while the second law states that energy becomes increasingly dispersed, as measured by the physical concept of entropy.

These laws are fundamental and seem somewhat hidden, but they can be experienced in everyday life. When I start my day with a hot cup of coffee, the inevitable will happen: the coffee will get colder. What actually happens is that the heat, or thermal energy, within the cup and my office are spread more and more evenly, which eventually results in equal temperatures (and cold coffee). During this redistribution, energy stays conserved (the first law), but the entropy increases (the second law). The increase in entropy caused by a process that spreads energy is a fundamental consequence of a process being active. This not only happens with cups of coffee, but with all processes in the Earth system. The increase of entropy sets a fundamental direction for anything that happens on Earth, and the speed at which it happens tells us about the overall activity. All processes follow these laws, the climate system, the biosphere, and human societies, with no exceptions.

There is more to the second law than just setting a direction. It sets limits to how much free energy, energy such as electric energy that is able to perform work and maintain activities, can be generated. This limit can be illustrated by a power plant (Figure 2). A power plant generates electric energy out of heat. The heat is generated by combustion of a fuel such as coal at a high temperature, which is energy of low entropy. The heat that leaves the cooling towers of the plant, seen in form of the white clouds emerging from them, exits the power plant at much colder temperature, hence at higher entropy. Overall, energy needs to be conserved, so the heat created by combustion balances the heat lost through the cooling towers and the electric energy being generated.

The second law tells us that not all of the heat from combustion can be converted into power that is fed into the electricity grid. A substantial fraction needs to leave the cooling towers, so that overall, at least as much entropy leaves the power plant as enters it by the combustion process. In the ideal case, these two rates are equal, which defines the well-known Carnot limit in thermodynamics. It is a direct consequence of the first and second law of thermodynamics. This limit of generation is set by the heat flux entering the power plant, and a fraction, $(T_{in} - T_{out})/T_{in}$, that is set by the difference in temperatures at which heat is added and removed. This fraction is also referred to as the Carnot efficiency.

This limit does not only apply to power plants, but also to solar panels (for this, one needs to use the entropy of radiation, e.g., Kabelac, 1994) and to how the Earth generates different forms of energy, for instance when it generates atmospheric motion out of heat. More generally, this limit sets an upper bound to how active a process can be.

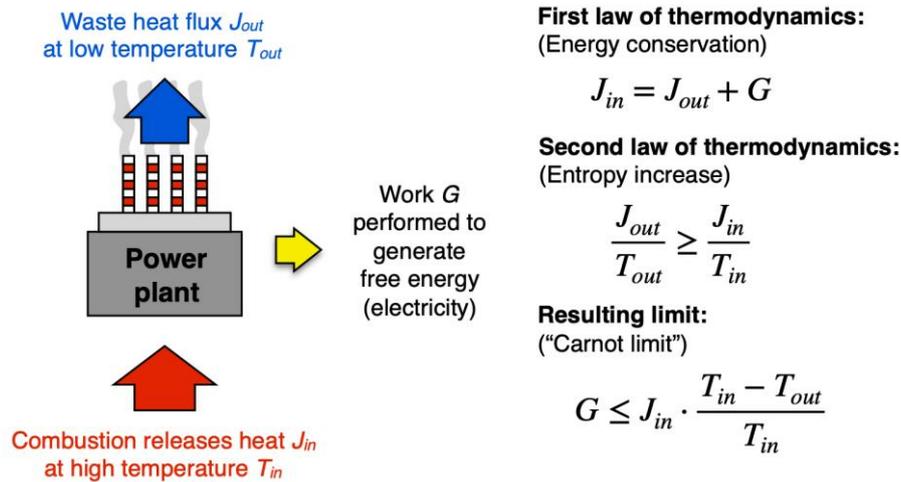

**Figure 2:** Illustration of the thermodynamic Carnot limit using a power plant as an example. The equations on the right show that this limit directly follows from the laws of thermodynamics.

**The Sun energises the Earth system**

When we apply the laws of thermodynamics to the Earth system, it is the continuous input of energy by sunlight that keeps the Earth active. Sunlight is radiation in the visible range of wavelengths. It has a very low entropy because it has been emitted by the Sun at a high temperature. After solar radiation has been absorbed, converted and redistributed within the Earth system, it is emitted to space as terrestrial radiation at much lower temperatures, having longer wavelengths in the infrared range and higher entropy. It is this difference in the entropy of the radiation the Earth receives and that it emits that allows for energy conversions and for the Earth to stay active (Figure 3).

Most of this difference in entropy between solar and terrestrial radiation is, however, destroyed when solar radiation turns into heat upon absorption at the Earth's surface. The difference in radiative heating at warmer places and radiative cooling at colder places can then be used by physical climate system processes to convert heat into physical forms of free energy. These energy conversions keep the climate system active, with organised circulation patterns in the atmosphere, ocean currents and hydrologic cycling.

The different forms of energy are generated by sunlight heating the Earth unevenly, thereby creating temperature differences. These temperature differences generate the kinetic energy of atmospheric motion, just like a power plant uses the temperature difference to generate electricity. The result of motion is that the heat it transports reduces these temperature differences, which lowers the ability to generate kinetic energy, or, in thermodynamic terms, it lowers the efficiency of the conversion process (Figure 4). Simple considerations demonstrate that the atmosphere operates near its thermodynamic limit of maximum power, generating about 1000 TW (1 TW = $10^{12}$ W) of kinetic energy. This is as much kinetic energy as it is possible to generate. In other words, the atmosphere works as hard as it can, and thus staying as active as possible. It suggests more generally that thermodynamic limits not just exist within the Earth system, but that they are highly relevant because physical processes actually evolve to and are maintained at these limits.

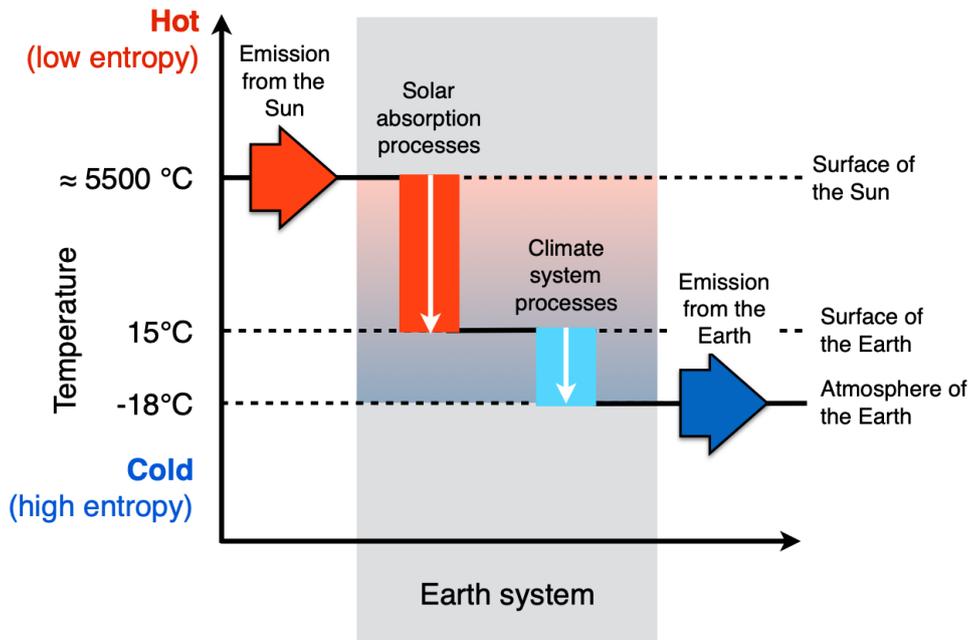

**Figure 3: The degradation of solar energy to higher entropy radiation as it is transformed by Earth system processes.**

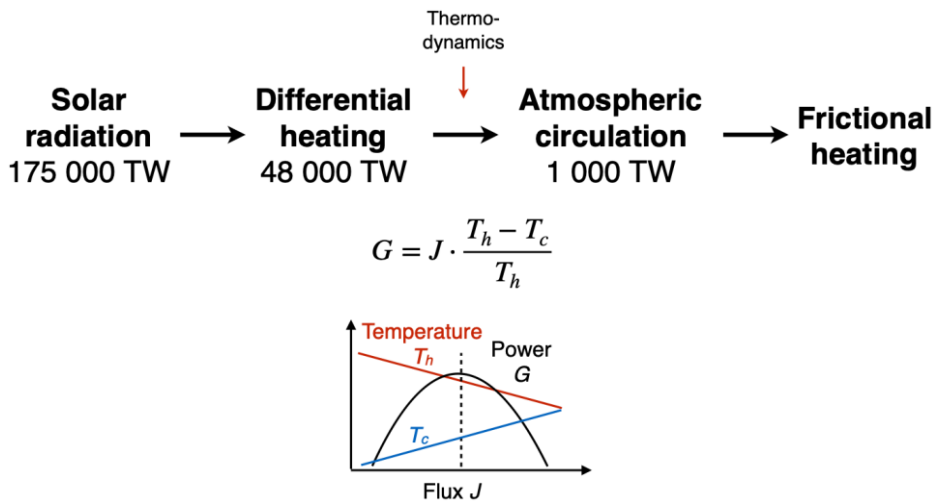

**Figure 4: Atmospheric motion results from uneven heating by absorption of solar radiation. The strength of the resulting atmospheric circulation, in terms of the rate by which kinetic energy is generated, is set by the thermodynamic limit and the effect that heat transport has on depleting the driving temperature difference.**

Motion, in turn, provides the means for further energy conversions. Upward motions cool air, bringing moisture to its saturation and causing precipitation, thus driving hydrologic cycling. The precipitated water on land dissolves minerals of the continental crust, thus driving geochemical weathering. Yet, the rates by which energy is converted and sunlight is utilised are comparatively small. The whole atmospheric circulation, for instance, operates by converting less than 1% of the incoming solar radiation into the kinetic energy associated with large-scale motion. This is because much of the potential of solar radiation to perform work is lost when it is absorbed and converted into heat. The resulting temperature differences on Earth are small compared to that of a power plant, so that relatively little power can be generated.

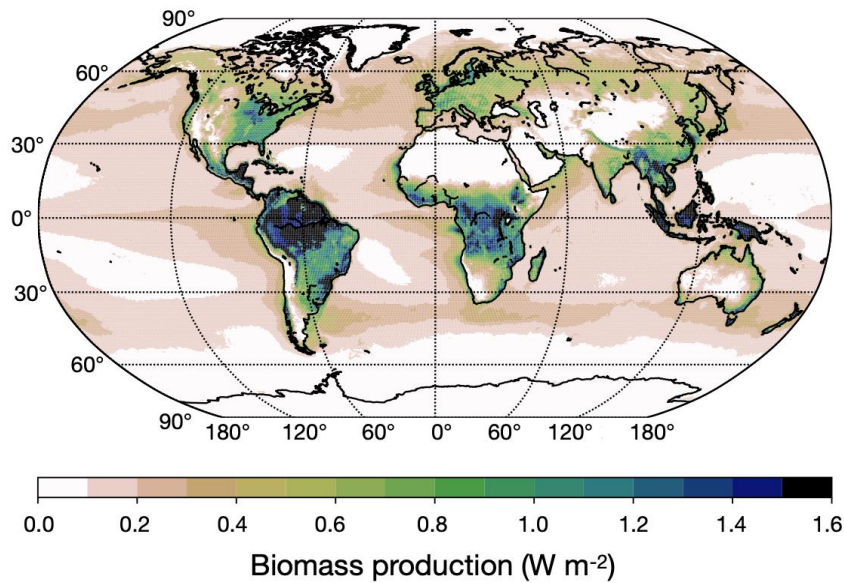

**Figure 5: Mean annual chemical free energy generation by the biosphere in form of biomass (net primary productivity, photosynthesis - autotrophic respiration), expressed in units of W m$^{-2}$. The map shows estimates derived from satellite and is taken from http://orca.science.oregonstate.edu/2160.by.4320.monthly.hdf.vgpm.m.chl.m.sst.php (ocean) and https://nacp-files.nacarbon.org/nacp-kawa-01/ (land). Accessed 19 December 2018.**

With photosynthesis, life created an innovation because it does not use heat and temperature differences as a means to generate energy, but it uses the low entropy of sunlight directly. It uses visible light to split water first into hydrogen and oxygen, and then the hydrogen further into its proton and electron. This generates electrical energy that is used to generate chemical energy in form of carbohydrates and atmospheric oxygen. Photosynthesis generates energy more efficiently than the physical climate system, yet it is nevertheless only able to convert less than 3% of the energy contained in sunlight. In many regions, biospheric activity is further limited by environmental factors (Figure 5). In the oceans, mixing limits the supply of nutrients, while on land, the presence of water is a major limitation. This leads to a strong imprint of climate on patterns of biological productivity, and it makes the ability of the biosphere to generate energy dependent upon the work done by the physical climate system. Overall, it leads to a generation of chemical free energy of about 220 TW.

The energy generated by photosynthesis feeds food webs of natural ecosystems, but also geochemical reactions in the Earth system. In fact, life produces substantially more chemical energy than abiotic processes, such as stratospheric ozone chemistry, nitrogen fixation by lighting, or geochemical weathering (Figure 6). Life thus plays a dominant role in shaping the geochemical composition of the planetary environment. This effect on the geochemical composition feeds back to the planetary system because the atmospheric composition affects how well radiation is being absorbed and emitted. This, in turn, alters heating and cooling rates and thereby the boundary conditions that determine how active the physical climate system can be. As climate limits biospheric activity, this results in a positive feedback by which life changes planetary geochemistry, which alters climate, which then allows the biosphere to become more active. This is probably the dominant feedback that shaped the evolution of the climate-biosphere system over the history of the Earth. It likely evolved the biosphere to become more and more active, which allowed it to become increasingly more complex and diverse (see also Judson, 2017).

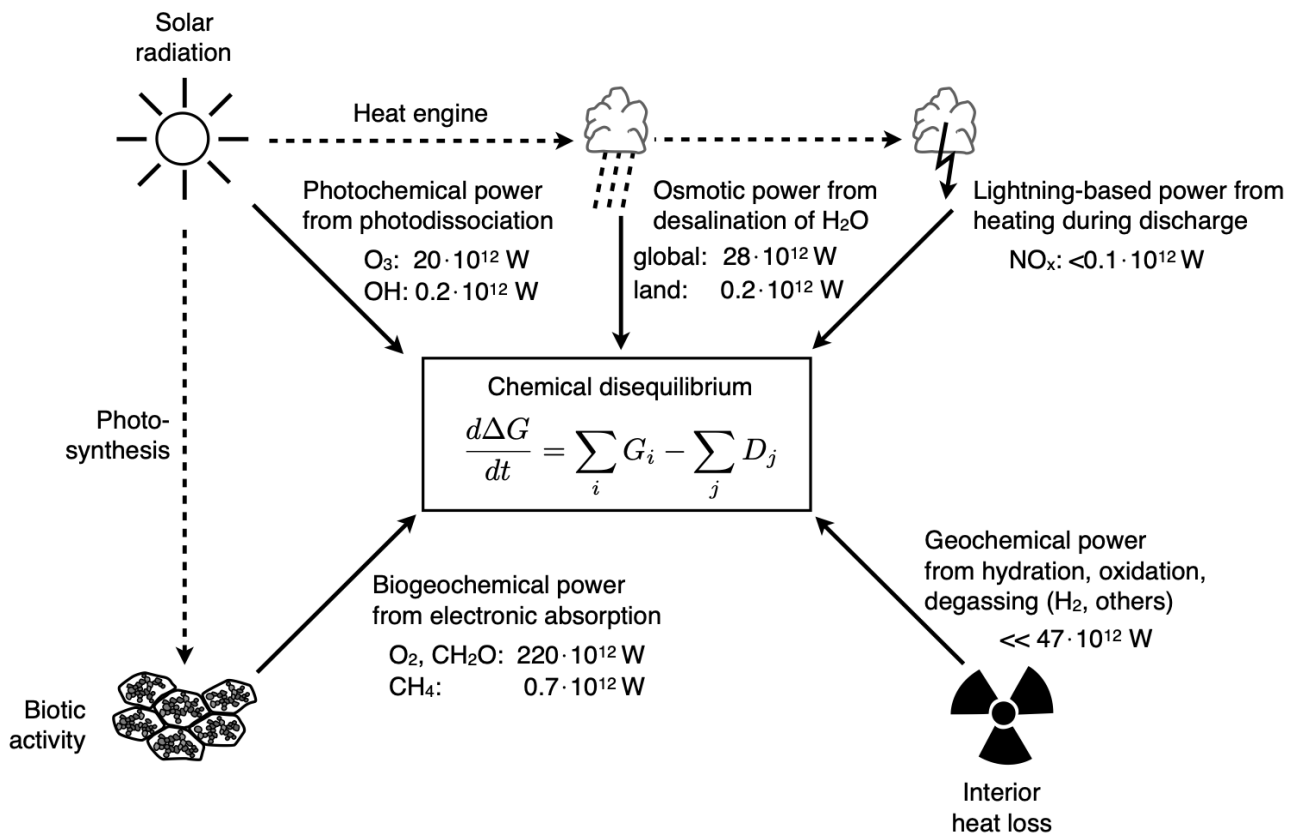

**Figure 6: Processes that generate chemical free energy in the Earth system (Kleidon, 2016).**

**4. Human activity as a thermodynamic Earth system process**

When we next turn to human activity in the Earth system, we start with the essential need for energy to maintain human metabolisms and human societies. The calories in food measure the energy that maintains human metabolisms. It comes from the harvests of the energy generated by photosynthesis from croplands and, indirectly, from livestock that grazes on pastures. The energy that fuels socioeconomic activity comes primarily from fossil fuels, stocks of energy contained in plant residues from millions of years ago that did not get decomposed because they were locked away by geologic processes.

To place these two forms of energy consumption by human activity into a quantitative context: The average human metabolism of a 75 kg human consumes about 100 W. Multiplied by 7.7 billion people, this yields a consumption rate of about 0.8 TW. It requires more energy from the biosphere to meet this demand because only a small fraction of plant tissues are suitable to be digested by humans. The use of animal products results in further inefficiencies as it needs to also sustain the metabolic activity of the livestock. Estimates place the actual use of photosynthesis for the metabolic needs of humans at about $8 \times 10^{12}$ W, which represents about 5% of the energy generated by photosynthesis on land (based on estimates by Haberl et al. 2014).

Socioeconomic activity is primarily driven by fossil fuels, or more generally, by primary energy consumption, which averages to a consumption rate of about $18 \times 10^{12}$ W within the year 2017 (BP, 2018). It is more than 20 times the metabolic energy needs of humans. The magnitude of this consumption rate emphasises how relevant human activity has become in the Earth system (Figure 1). Socioeconomic activity consumes more energy than, for instance, the power that mixes the world's oceans, which is about $5 \times 10^{12}$ W. It substantiates the notion that we entered the geologic era of the Anthropocene (Crutzen, 2002). Human activity and the associated socioeconomic activity consume energy at rates similar to or even greater than those of natural processes of the Earth system.

From the thermodynamic perspective of the Earth system, the consequences of human activity are fairly straightforward. Human activity consumes forms of free energy that was generated by other Earth system processes, so it depletes associated states of thermodynamic disequilibrium. And this depletion of disequilibrium states is linked with many of the problems of global change.

Using an increasing share of the plant productivity of the biosphere as the energy source to feed the human population means that less energy is available to sustain natural ecosystems and their food chains. The consequences manifest themselves in increased land cover change, reduced wildlife, deteriorated ecosystems and biodiversity loss.

Using fossil fuels as the primary energy source for socioeconomic activity depletes the geochemical disequilibrium between hydrocarbons in geologic reservoirs and atmospheric oxygen. This major form of disequilibrium of the Earth system was generated by the Earth's biosphere and geologic processes over millions of years in the past. As a result of digging up fossil fuels and burning them, carbon dioxide levels in the Earth's atmosphere have increased, causing an enhanced greenhouse effect, global warming, and associated changes in the climate system. A cool climate with a carefully regulated greenhouse gas composition does not come for free — it requires chemical work to actively remove carbon dioxide from the atmosphere. This is what the biosphere has done over the Earth's history. Human activities at present undo this work by consuming fossil fuels and by burning up the associated form of disequilibrium.

Human activity is thus, overall, a planetary process that, measured by its magnitude of energy consumption, compares to other planetary processes. Yet, it only consumes energy, resulting in detrimental effects that deplete the planetary state of disequilibrium, making Earth less active.

## 5. Imagining a sustainable future

Let us now get back to the original motivation and the question how one can imagine a sustainable future from this thermodynamic perspective. Energy consumption is deeply engrained in socioeconomic activities. It is needed for heating and cooling, to build and maintain infrastructures, to transport and trade resources and goods, and transcends most human activities. While increases in efficiency can reduce human energy consumption to some extent, a forced reduction in energy consumption is likely to have devastating consequences for human societies.

> **Energy consumption is deeply engrained in socioeconomic activities.**

So if the current level of energy consumption is going to be more or less maintained or needs to grow further as the world population and standards of living rise, how can this demand be sustainably met? By its magnitude, it compares with other Earth system processes, so how can that much free energy be made available to human societies without detrimental effects on the Earth system?

The answer to this question is human-made technology, specifically photovoltaics. Just as photosynthesis, it uses sunlight directly to perform the work of charge separation to generate electric energy. In this respect, it seems to achieve the same goal. Yet, in contrast to photosynthesis, a solar panel does not rely on sources of carbon, water and nutrients as it exports its free energy solely in electric form. Even current, industrial-grade solar panels are vastly more efficient than photosynthesis, using about 20% of the solar radiation, rather than less than 3% of the solar radiation that photosynthesis can utilise. The theoretical limits of using sunlight of more than 70% will allow for further technological advances and even higher efficiencies. What this signifies is that humans created a major innovation by developing a more efficient process of getting free energy from sunlight than what abiotic or biotic processes have been able to accomplish so far.

> In contrast to photosynthesis, a solar panel does not rely on sources of carbon, water and nutrients as it exports its free energy solely in electric form.

To put this form of sustainable energy generation in numbers: To generate the current primary energy consumption of about 18 TW, it would require solar panels with an efficiency of 20% covering about 400 000 km$^2$ of barren land (about the size of Germany). Given that barren lands cover 19 million km$^2$ of Earth's continental areas, this represents a small fraction of the world's deserts. On this land, solar radiation would not be wasted by turning into heat upon absorption, but would generate free energy to sustain human societies. In other words, it would allow the whole Earth system to produce more free energy.

It would then require little effort to generate even more energy by expanding the use of photovoltaics. This additional energy could be used for desalination to provide more freshwater, more than the natural hydrologic cycle can provide. For desalinating seawater, evaporation requires about 2.5 MJ for each litre of seawater, which then provides freshwater when it rains out. Desalination technology uses membranes, and so the energy requirement to desalinate seawater shrinks to merely 6.5 kJ for each litre of seawater (Elimelech and Phillip, 2011). In other words, with human-made technology, one can enhance the hydrologic cycle by having more efficient options for desalination than how natural processes generate freshwater.

> With human-made technology, one can enhance the hydrologic cycle.

This additional energy and water, made available by human technologies, could then be used to expand agriculture — either into desert regions, or in form of vertical farming, which grows crops in warehouses on multiple shelves with artificial lighting and closed water- and nutrient cycling. This would take food production to unprecedented levels, as it can take agricultural production out of the naturally constrained environment. This form of technology-based agricultural expansion would reduce the pressure to clear and convert productive natural ecosystems, such as tropical rainforests, to be used for this expansion. It would even allow to increase overall productivity, thus enlarging Earth's biosphere.

With the heavy use of human technology, the Earth would clearly start to look different from its natural state. Yet, it would, overall, be an evolution in which human technology provides means to enhance the free energy generation of the planet, so humans become producers of free energy, rather than merely consume what the Earth has generated. The human species alone cannot accomplish this task because by nature, humans are consumers of energy, not producers. It needs solar-based technology that is more efficient than natural photosynthesis for this evolution to take place. Overall, it would allow the Earth to become a more active, dissipative system.

## 6. Conclusions

What I described here is a quantification of the human role in the Earth system that is based on thermodynamics and free energy, energy that is able to perform work. It shows the huge role that humans play in terms of the energy they consume, which is similar in magnitude to many natural processes of the Earth system. It relates many of the detrimental impacts of human activity on the planet to the consumption of energy that the Earth system has generated, so that a sustainable future with further growth can be envisioned if human societies turn into producers of free energy by the use of technology. With the use of technologies such as photovoltaics, humans can produce energy more efficiently than natural processes, and this allows for stronger material cycling, for instance in terms of stronger hydrologic cycle, enhanced agricultural production, or enhanced industrial activity. In this sense, through human-made technology, the Earth can, as a whole, become a more active, dissipative system.

> **Human societies should turn into producers of free energy by the use of technology.**

What is so easily described here is obviously not a simple task to be accomplished. It would require human societies to make rational choices made on objective, physical reasoning. It would further require to think beyond simplistic cause-and-effect lines of thought, as systems are strongly shaped by interactions, which in turn affect the thermodynamic state of the planet. These interactions do not necessarily manifest themselves immediately, so it also requires a long-term perspective that includes interactions at various scales. In other words, it would require us to think about the consequences in the context of the thermodynamic state of the whole Earth system. Last, but not least, it would require us to accept a certain level of change as well as the associated responsibility for that change, as the Earth system would unavoidably evolve towards an era of human domination.

> **Among several other factors human societies are to to make rational choices made on objective, physical reasoning.**